\documentclass[12pt,manuscript]{aastex}
\usepackage{amsmath, amsfonts, color}
\usepackage{placeins, mathrsfs, natbib}
\def\ltsima{$\; \buildrel < \over \sim \;$}
\def\ltsim{\lower.5ex\hbox{\ltsima}}
\def\gtsima{$\; \buildrel > \over \sim \;$}
\def\gtsim{\lower.5ex\hbox{\gtsima}}

\newcommand{\paren}[1]{\left(#1\right)}
\newcommand{\brak}[1]{\left[#1\right]}

\newcommand{\lparen}[1]{\left(#1\right.}
\newcommand{\rparen}[1]{\left.#1\right)}

\newcommand{\lbraket}[1]{\left<#1\right.}
\newcommand{\rbraket}[1]{\left.#1\right>}

\newcommand{\abs}[1]{\left|#1\right|}
\newcommand{\braket}[1]{\left<#1\right>}

\newcommand{\phivec}{\hat{\bs{\phi}}}
\newcommand{\Rvec}{\hat{\bs{R}}}
\newcommand{\zvec}{\hat{\bs{z}}}

\newcommand{\bs}[1]{\boldsymbol{#1}}

\newcommand{\parti}[2]{\frac{\partial#1}{\partial#2}}

\newcommand{\di}[2]{\frac{d#1}{d#2}}

\begin{document}
\title{The Magnetoviscous-thermal Instability}
\author{Tanim Islam}
\affil{Lawrence Livermore National Laboratory, P.~O.~Box 808, Livermore, CA 94551-0808}
\email{islam5@llnl.gov}

\begin{abstract}
Accretion flows onto underluminous black holes, such as Sagittarius A* at the center of our galaxy, are dilute (mildly collisional to highly collisionless), optically thin, and radiatively inefficient. Therefore, the accretion properties of such dilute flows are expected to be modified by their large viscosities and thermal conductivities. Second, turbulence within these systems needs to transport angular momentum as well as thermal energy generated through gravitational infall outwards in order to allow accretion to occur. This is in contrast to classical accretion flows, in which the energy generated through accretion down a gravitational well is locally radiated. In this paper, using an incompressible fluid treatment of an ionized gas, we expand on previous research by considering the stability properties of a magnetized rotating plasma wherein the thermal conductivity and viscosity are not negligible and may be dynamically important. We find a class of MHD instabilities that can transport angular momentum and thermal energy outwards. They are plausible candidates to describe accretion in radiatively inefficient accretion flows (RIAFs). We finish by discussing the implications for analytic models  and numerical MHD simulations of mildly dilute or collisionless astrophysical plasmas, and immediate directions for further research.
\end{abstract}

\maketitle

\section{Introduction}
The study of astrophysical rotating plasmas in general, and accreting plasmas in particular, has undergone a renaissance within the past twenty years. \citet{Balbus91, Hawley91} reapplied the magnetorotational instability (MRI, \citet{Velikhov59, Chandrasekhar60}) into an appropriate astrophysical context. They demonstrated that in rotating astrophysical plasmas, even arbitrarily small magnetic forces can provide a channel by which free energy source of outwardly decreasing angular velocity, rather than angular momentum, can destabilize a Rayleigh-stable flow, generate magnetic fields with energies of order the thermal energy, and drive the right form of turbulence that allows for accretion to occur. Numerous numerical simulations \citep{Hawley96, Wardle99, Sano02, DeVilliers03a, Fromang04c} have borne out the fact that magnetic fields can play an essential role in driving efficient accretion for a wide class of astrophysical objects. The nonlinear MRI is currently the most plausible candidate to explain high mass-rate, radiatively efficient accretion onto massive and supermassive central galactic black holes ($M \gtsim 10^7 M_{\odot}$), which manifests itself as active galactic nuclei (AGNs) and quasars.

Although AGNs and quasars are some of the most powerful and most energetic compact objects in the universe, at recent epochs in the history of the universe (at redshifts less than 1) they are rare. Evidence from Very Large Baseline Interferometry (VLBI) measurements of water masers, from gas spectroscopy, and from stellar dynamics give strong evidence for the existence of massive and supermassive black holes in galaxies that possess a central bulge (see \citet{Richstone98} for an exhaustive review on diagnostic tools for the detection of central galactic black holes). However, recent X-ray surveys of galactic and extragalactic sources \citep{Tueller08, Tueller10} has demonstrated that there is strong observational evidence of approximately 230 local AGNs, located within  Seyfert galaxies, within 400 million light-years. The relative transparency of active galactic nuclei hard X-ray emission to the high column density material obscuring them \citep{Treister05}, and the lack of confusion with other sources of emission in the AGN’s host galaxy \citep{Markwardt05}, makes hard X-ray emission an optimal spectral regime for unbiased AGN searches. Within this 400 million light-year volume of space are at least $\sim 10^4$ galaxies that possess a central bulge. Only a few percent of central galactic black holes at the current epoch are AGNs or quasars. It is therefore plausible that most accretion onto massive and supermassive central galactic black holes systems at the current epoch is highly radiatively inefficient and dilute.

Although the MRI is one of the more well-known free-energy gradient instabilities with astrophysical applications, in dilute astrophysical plasmas, such as radiatively inefficient accretion flows (RIAFs), large viscosities \citep{Balbus04b, Islam2005}, and thermal conductivities \citep{Balbus01a} can also destabilize plasmas that possess free energy gradients. This unusual behavior arises from the fact that even a magnetic field with dynamically unimportant Lorentz forces can easily be strong enough that the ion Larmor radius is smaller than the collisional mean free path. Under these conditions, the viscous stress and thermal flux are dynamically important and directed along magnetic field lines \citep{Braginskii65}.

The explanations for the magnetothermal instability (MTI) and magnetoviscous instability (MVI) are conceptually simple. An equilibrium is perturbed so that there is a component of field line, hence of thermal conductivity (viscosity), along the thermal (angular velocity) gradient; heat (momentum) can then be transferred outwards, which further deforms magnetic field lines along the direction of the gradient -- the process runs away. Similar behavior arises in the collisionless regime, in which large heat fluxes (momentum fluxes) along magnetic fields act to destabilize the plasma. This phenomenon is shown in collisionless and mildly collisional treatments of the MRI \citep{Quataert02c, Sharma03}. Furthermore, consistent collisionless MHD simulations \citep{Sharma06a, Sharma07b} incorporate the fact that the arbitrary development of pressure anisotropy is limited by pitch-angle scattering due to fast gyrokinetic instabilities that violate magnetic adiabatic invariance for electrons and ions \citep{Gary94b, Gary96b, Gary98}.

Numerical simulations have also begun to demonstrate how large anisotropic transport coefficients can destabilize a plasma that contains free energy gradients. More recent, comprehensive, local and global nonlinear simulations of the MTI in a standard MTI-unstable simulation box \citep{McCourt11} demonstrated steady states characterized by regions with relatively tangled magnetic field configurations, developed kinetic turbulence, driven by the free energy of an upwardly decreasing temperature profile. Furthermore, numerical nonlinear simulations of the collisionless MRI \citep{Sharma06a, Sharma06}, which allows for efficient transport of angular momentum along field lines, also demonstrate the development of outwardly angular-momentum transporting turbulence, to a nearly steady state, in a local Keplerian rotational box (MRI-unstable, in that angular velocity decreases radially outwards).

This paper expands on previous research by considering an incompressible fluid treatment of instability in a rotating plasma with large anisotropic ion viscous stresses as well as electron thermal conductivities. This is referred to as the magnetoviscous-thermal instability (MVTI). The organization of this paper is as follows. In \S 2 we demonstrate that the MVTI may play an important role in describing the dynamics of accretion flows about dim galactic nuclei, taking Sagittarius A* as an example. In \S 3 we lay down the equations of the system, describe our equilibrium state, and derive expressions for angular momentum and total energy balance in dilute, radiatively inefficient magnetized accretion flows for thin, largely rotationally supported disks. In \S 4 we derive the dispersion relation. In \S 5 we justify and derive bulk fluxes for the MVTI, demonstrating that turbulence driven by this MHD instability can drive accretion onto radiatively inefficient flows. Finally, in \S 6 we provide a short discussion of the conclusions, its main astrophysical implications, as well as directions for further research; a companion paper will explore the MVTI in the collisionless limit, and the corresponding instability will be termed the collisionless MTI.

\section{Preliminaries}
We briefly review the conditions of validity for the approximations used in this calculation, and note the forms of viscous stress and heat flux in a magnetized fluid. First, we note that a fluid approach for a low density plasma requires that the ion and electron mean free paths be small compared to global length scales.  In our calculation, the ions and electrons are assumed to have the same temperature, so the ion-ion and electron-electron mean free paths are identical. For convenience, however, we will always refer to the ion mean free path $\lambda_i$,
\begin{eqnarray}
  &&\lambda_i = 1.5\times 10^{13} \frac{T_4^2}{n_1\ln\Lambda}\text{ cm},
\end{eqnarray}
where $T_4$ is the temperature in units of $10^4$ K, $n_1$ is the number density in cm$^{-3}$, and $\ln\Lambda$ is the Coulomb logarithm. The second condition is that this mean free path be large compared to the ion Larmor radius $r_L$,
\begin{eqnarray}
  &&r_L = 9.5\times 10^7 B_{\mu G}^{-1} T_4^{1/2}\text{ cm},
\end{eqnarray}
so that transport is directed along magnetic lines of force. Here $B_{\mu G}$ is the magnetic field strength in $\mu$G. The electron Larmor radius is of course a factor of 43 smaller than the ion Larmor radius.

The formal regime of validity is $r_L \ll \lambda_i \ll R$, where $R$ is a typical global scale of the flow.  In practice the lower bound on $\lambda_i$ is expected to be well-satisfied in the applications of interest (the protogalactic interstellar medium and under luminous black hole accretion).  The upper bound on $\lambda_i$, on the other hand, breaks down in the innermost regions of black hole accretion models for the Galactic Center \citep{Quataert04}. Our fluid approximation is expected to be well-satisfied throughout the bulk of this flow, from regions of a few times $10^4$ Schwarzchild radii out to Sag.~A*'s Bondi capture radius and the protogalactic interstellar medium.

\section{Formulation of the Problem}
Dim mass-starved accretion flows onto black holes are radiatively inefficient. Under these circumstances, the turbulent energy flux should contain a substantial purely thermal contribution (from correlated fluctuations in the density and temperature) as well the usual contribution from the angular momentum transport \citep{Balbus03}. Because of its role in redistributing energy, this additional thermal flux could have a profound effect on the observational properties of the flow. It is the possibility that the MVTI could be the source of a profoundly enhanced thermal energy flux that motivates our study, beginning with an elucidation of its linear properties.

We examine the rotational and convective stability of a disk under the effects of a magnetic field, under the effects of a large Braginskii viscosity \citep{Balbus04b, Islam2005} as well as large thermal conductivity \citep{Balbus01a}. Our coordinate system for the disk will be a standard cylindrical system: radius $R$, azimuth $\phi$, and axial variable $z$. As in \citet{Balbus01a}, we consider the stability of an equilibrium plasma at the midplane with only radial gradients in pressure and temperature, and nonradial equilibrium magnetic field ${\bf B}_0 = B_0\paren{\phivec\cos\chi\phivec + \zvec\sin\chi}$, where $B_0$ is the magnitude of the equilibrium magnetic field. We analyze unstable axisymmetric modes in the Boussinesq limit -- incompressible flow and isobaric perturbations. We calculate the linear stability of an idealized plasma with a single temperature $T$ with physical radially outwardly decreasing temperature and pressure that remains convectively stable, and demonstrate that its dispersion relation reduces to that of the MRI, MVI, and MTI in specific limiting cases. We also demonstrate that the quadratic heat fluxes and Reynolds stresses associated with the MVTI are of the right sense to drive accretion in radiatively inefficient astrophysical plasmas.

\subsection{Constituent Equations}
The fluid equations consist of continuity, force balance, energy balance, and the induction equations:
\begin{eqnarray}
  &&\parti{\rho}{t} + \nabla\cdot\paren{\rho{\bf V}} = 0, \label{eq:cont} \\
  &&\rho\paren{\parti{\bf V}{t} + {\bf V}\cdot\nabla{\bf v}} = -\nabla\paren{p + \frac{{\bf B}\cdot{\bf B}}{4\pi}} + \frac{{\bf B}\cdot\nabla{\bf B}}{4\pi} - \rho\nabla\Phi - \nabla\cdot\bs{\sigma}, \label{eq:forcebalance} \\
  &&\frac{3}{2} p\di{\ln p\rho^{-5/3}}{t} = -\nabla\cdot\paren{q{\bf b}} - \bs{\sigma} : \nabla{\bf V}, \label{eq:energybalance} \\
  &&\parti{\bf B}{t} = \nabla\times\paren{{\bf V}\times{\bf B}}, \label{eq:induction}
\end{eqnarray}
where $\rho$ is density; ${\bf V}$ is the velocity; $\Phi$ is the gravitational potential; ${\bf B}$ is the magnetic field; $\bs{\sigma}$ is the viscous stress tensor; $q$ is the anisotropic heat flux; and $P$ is the pressure. ${\bf b} = {\bf B}/B$ is the unit vector along the magnetic field line. The viscous stress tensor and heat flux are given in \citet{Braginskii65}:
\begin{eqnarray}
  &&\bs{\sigma} = -3\rho\nu \paren{{\bf b}{\bf b} - \frac{1}{3}\mathbb{I}}\paren{{\bf b}\cdot\nabla{\bf V}\cdot{\bf b} - \frac{1}{3}\nabla\cdot{\bf V}}, \label{eq:sigma_total} \\
  &&q = -\kappa\rho {\bf b}\cdot\nabla\theta \label{eq:q_total},
\end{eqnarray}
where $\nu$ and $\kappa$ are the viscous and (electron) thermal diffusivity along the magnetic field line, $\theta = p/\rho = k_B T/m_i$ is the isothermal sound speed squared, and $m_i$ is the ion mass.

\subsection{Fluxes in Dilute, Magnetized Disks}
Here we derive expressions for the heat and angular momentum fluxes in geometrically thin, dilute, radiatively inefficient, magnetized disks. Since velocity fluctuations are possibly of the order of the disk sound speed, then the dominant contributions to turbulence that carries out angular momentum and (as we shall see) thermal energy to allow accretion is through quadratic correlations between fluctuating fluid quantities. We first find it useful to transform to a frame corotating with the equilibrium flow ${\bf V}_0 = R\Omega(R)\phivec$, and define a flow velocity ${\bf v} = {\bf V} - {\bf V}_0$ with respect to the equilibrium. The equilibrium force balance equation is given by,
\begin{eqnarray}
  &&-R^2\Omega(R)\Rvec = -\frac{1}{\rho_0}\nabla p_0 - \frac{G M}{R^2}\Rvec - \frac{G M}{R^3}z\zvec. \label{eq:equilibrium_force_balance}
\end{eqnarray}
If we employ a variable transformation ${\bf V} \to {\bf v} = {\bf V} - R\Omega(R)\phivec$, substitute in the forms of the viscous stress tensor (Eq.~[\ref{eq:sigma_total}]) and heat flux (Eq.~[\ref{eq:q_total}]), and take into account equilibrium force balance (Eq.~[\ref{eq:equilibrium_force_balance}]), then the dynamic MHD equations, Eqs.~(\ref{eq:cont}) - (\ref{eq:induction}), become,
\begin{eqnarray}
  &&\paren{\parti{}{t} + \Omega\parti{}{\phi}}\rho + \nabla\cdot\paren{\rho{\bf v}} = 0, \label{eq:continuity_comoving} \\
  &&\begin{aligned}{}
      &\rho\paren{\brak{\parti{}{t} + \Omega\parti{}{\phi}} + {\bf v}\cdot\nabla{\bf v} - 2\Omega{\bf v}\times\zvec + \Omega'R u_R \phivec}{\bf v} = -\nabla\paren{p + \frac{B^2}{8\pi}} + \\
      &\frac{{\bf B}\cdot\nabla{\bf B}}{4\pi} + \frac{\rho}{\rho_0}\nabla p_0 - \nabla\cdot\bs{\sigma},
    \end{aligned} \label{eq:force_balance_comoving} \\
  &&\frac{3}{2}p\paren{\parti{}{t} + \Omega\parti{}{\phi}}\ln p\rho^{-5/3} + \frac{3}{2}p{\bf v}\cdot\nabla \ln p \rho^{-5/3} = \nabla\cdot\paren{q{\bf b}} - \bs{\sigma} : \nabla{\bf v}, \label{eq:MHD.internal_energy_balance_comoving} \\
  &&\paren{\parti{}{t} + \Omega\parti{}{\phi}}{\bf B} + {\bf v}\cdot\nabla{\bf B} = -{\bf B}\paren{\nabla\cdot{\bf v}} + {\bf B}\cdot\nabla{\bf v} + \Omega'R B_R \phivec. \label{eq:MHD.induction_comoving}
\end{eqnarray}
We use a prescription for averaging out turbulent fluctuations as laid out in \citet{Balbus98}. We assume a disk that is steady and spatially smooth over time and length scales much larger than the turbulence. The fluctuation in fluid velocities $\paren{u_{\phi}, u_z}$, magnetic field ${\bf B}$, and density $\rho$ average to zero.
\begin{eqnarray}
  &&\braket{\delta u_{\phi}}_{\rho} = \braket{\delta u_z}_{\rho} = 0, \label{eq:fluctuation_velocity_to_zero} \\
  &&\braket{\delta{\bf B}}_{\rho} = {\bf 0}, \label{eq:fluctuation_Bfield_to_zero} \\
  &&\braket{\delta\rho}_{\rho} = 0. \label{eq:fluctuation_density_to_zero}
\end{eqnarray}
This average, $\braket{}_{\rho}$, is defined in the following manner:
\begin{eqnarray}
  &&\braket{A}_{\rho} = \frac{1}{2\pi\Sigma\Delta R \Delta T} \int_0^{\Delta T} \int_{R-\Delta R/2}^{R+\Delta R/2} \int_{\phi = 0}^{2\pi} \int_{z=-\infty}^{\infty} A \rho\,dR'\,dz\,d\phi\,dt \label{eq:accretionmodel.averaged_quantity},
\end{eqnarray}
where,
\begin{eqnarray}
  &&\Sigma = \int_{z=-\infty}^{\infty} \rho\,dz \label{eq:accretionmodel.surface_mass_density}.
\end{eqnarray}
This average is taken over a radial slice $\Delta R$ much smaller than the disk radius but larger than the largest scale of the turbulence and a time scale $\Delta T$ much longer than the turnover timescale of the turbulence. The radial velocity $u_R$ consists of a fluctuating part $\braket{\delta u_R}_{\rho} = 0$ and a nonzero mean mass flow. The mean flow velocity is far smaller than the root mean square turbulent velocity, i.e., $\abs{\braket{u_R}_{\rho}} \ll \sqrt{\abs{\braket{\delta u_R \delta u_{\phi}}_{\rho}}}$.

The angular momentum density in the $z$-direction, ${\mathcal L}_z = \rho R V_{\phi}$, can be found from Eqs.~(\ref{eq:cont}) and (\ref{eq:forcebalance}). After some algebra, we have,
\begin{eqnarray}
  &&\parti{{\mathcal L}_z}{t} + \nabla\cdot\paren{R\brak{\rho V_{\phi}{\bf V} - \frac{B_{\phi}{\bf B}}{4\pi} + \paren{p + \frac{B^2}{8\pi}}\phivec + \bs{\sigma}\cdot\phivec}} = 0. \label{eq:angular_momentum_balance_intermediate}
\end{eqnarray}
If we substitute ${\bf v}$ for ${\bf V}$ and average according to the prescription given by Eq.~(\ref{eq:accretionmodel.averaged_quantity}), angular momentum balance is given by,
\begin{eqnarray}
  &&\parti{\braket{{\mathcal L}_z}_{\rho}}{t} + \frac{1}{R}\parti{}{R}\paren{R^2\braket{T_{R\phi}} + R^3\Omega\braket{\rho u_R}_{\rho}} = 0 \label{eq:dilute_angmom_average}.
\end{eqnarray}
The averaged quadratic angular momentum flux is,
\begin{eqnarray}
  &&\begin{aligned}{}
      &\braket{T_{R\phi}} = \lbraket{\rho_0 \delta u_R \delta u_{\phi} - \frac{\delta B_R \delta B_{\phi}}{4\pi}} - \\
      &\rbraket{3\rho_0 \nu \brak{{\bf b}_0\cdot\nabla\delta{\bf v}\cdot{\bf b}_0 + \Omega'R \delta b_R \cos\chi}\times \delta b_R \cos\chi\rule{0em}{1.5em}}_{\rho}.
    \end{aligned} \label{eq:averaged_angular_momentum_flux}
\end{eqnarray}

The local total energy balance equation can be found in steps. First, if we dot Eq.~(\ref{eq:force_balance_comoving}) with ${\bf v}$, and use Eq.~(\ref{eq:continuity_comoving}), we have,
\begin{eqnarray}
  &&\begin{aligned}
      &{\bf v}\cdot\paren{\parti{}{t} + \Omega\parti{}{\phi}}\paren{\rho{\bf v}} + {\bf v}\cdot\nabla p + u_i \partial_j\paren{\rho u_i u_j} + \\
      &u_i \partial_j\paren{-3\rho \nu\brak{{\bf b}\cdot\nabla{\bf v}\cdot{\bf b} - \frac{1}{3}\nabla\cdot{\bf v}}\brak{b_i b_j - \frac{1}{3}\delta_{ij}}} + \Omega'R \rho u_R u_{\phi} = \\
      &\frac{\rho}{\rho_0}{\bf v}\cdot\nabla p_0 - {\bf v}\cdot\frac{\paren{\nabla \times{\bf B}}\times{\bf B}}{4\pi}
    \end{aligned} \label{eq:MHD.mechanical_energy_balance}
\end{eqnarray}
Second, if we dot Eq.~(\ref{eq:MHD.induction_comoving}) we get the equation for the magnetic energy density evolution,
\begin{eqnarray}
  &&\begin{aligned}
      &\paren{\parti{}{t} + \Omega\parti{}{\phi}}\frac{B^2}{8\pi} = -{\bf v}\cdot\nabla\paren{\frac{B^2}{8\pi}} - \frac{B^2}{4\pi}\nabla\cdot{\bf v} + \frac{B^2}{4\pi}{\bf b}\cdot\nabla{\bf v}\cdot{\bf b} + \\
      &\Omega'R \frac{B_R B_{\phi}}{4\pi}.
    \end{aligned} \label{eq:MHD.magnetic_energy_equation}
\end{eqnarray}
Third, the MHD comoving internal energy equation, Eq.~(\ref{eq:MHD.internal_energy_balance_comoving}), is given by,
\begin{eqnarray}
  &&\begin{aligned}
      &\frac{3}{2}\paren{\parti{}{t} + \Omega\parti{}{\phi}}p + \nabla\cdot\paren{\frac{3}{2}p{\bf v}} + p\nabla\cdot{\bf v} = \nabla\cdot\paren{\kappa\rho\brak{{\bf b}\cdot\nabla\theta}{\bf b}} + \\
      &3\rho\nu\paren{{\bf b}\cdot\nabla{\bf v}\cdot{\bf b} - \frac{1}{3}\nabla\cdot{\bf v}}\paren{{\bf b}\cdot\nabla{\bf v}\cdot{\bf b} + R{\bf b}\cdot\nabla\Omega b_{\phi} - \frac{1}{3}\nabla\cdot{\bf v}}.
    \end{aligned} \label{eq:MHD.internal_energy_equation}
\end{eqnarray}
We add Eqs~(\ref{eq:MHD.mechanical_energy_balance}), (\ref{eq:MHD.magnetic_energy_equation}), and (\ref{eq:MHD.internal_energy_equation}), we arrive at the total energy balance equation,
\begin{eqnarray}
  &&\begin{aligned}
      &\paren{\parti{}{t} + \Omega\parti{}{\phi}}\paren{\frac{1}{2} \rho u^2 + \frac{3}{2} p + \frac{B^2}{8\pi}} + \nabla\cdot\lparen{{\bf v}\brak{\frac{1}{2}\rho u^2 + \frac{5}{2}p} + \frac{{\bf B}\times\paren{{\bf v}\times{\bf B}}}{4\pi} -} \\
      &\rparen{3\rho\nu\brak{{\bf b}\cdot\nabla{\bf v}\cdot{\bf b} - \frac{1}{3}\nabla\cdot{\bf v}}\brak{{\bf b}\paren{{\bf v}\cdot{\bf b}} - \frac{1}{3}{\bf v}}} - \rho{\bf v}\cdot\frac{1}{\rho_0}\nabla p_0 = \\
      &-\parti{\Omega}{\ln R}\paren{\rho u_R u_{\phi} - \frac{B_R B_{\phi}}{4\pi} - 3\rho\nu\brak{{\bf b}\cdot\nabla{\bf v}\cdot{\bf b} + R{\bf b}\cdot\nabla\Omega b_{\phi} - \frac{1}{3}\nabla\cdot{\bf v}}b_R b_{\phi}}.
    \end{aligned} \label{eq:MHD.total_energy_balance_unsimplified}
\end{eqnarray}
Applying the prescription for the average as given by Eq.~(\ref{eq:accretionmodel.averaged_quantity}) to Eq.~(\ref{eq:MHD.total_energy_balance_unsimplified}), we arrive at the formula for the averaged total energy balance equation,
\begin{eqnarray}
  &&\parti{\braket{\mathcal E}}{t} + \frac{1}{R}\parti{}{R} R\braket{F_{ER}} - \braket{\rho u_R}_{\rho} \frac{1}{\rho_0}\parti{p_0}{R} = -\parti{\Omega}{\ln R}\braket{T_{R\phi}}, \label{eq:MHD.dilute_energy_balance}
\end{eqnarray}
where,
\begin{eqnarray}
  &&\braket{\mathcal E} = \braket{\frac{1}{2}\rho u^2 + \frac{3}{2}p + \frac{B^2}{8\pi}}_{\rho}, \label{eq:MHD.total_energy_averaged} \\
  &&\begin{aligned}{}
      &\braket{F_{ER}} = \frac{5}{2}\rho_0 \braket{\delta u_R \delta\theta}_{\rho} - \kappa \rho_0\braket{\delta b_R \delta b_R \parti{\theta_0}{R}}_{\rho} + \\
      &\rho_0 \nu \braket{\paren{{\bf b}_0\cdot\nabla\delta{\bf v}\cdot{\bf b}_0 + \Omega'R \delta b_R \cos\chi} \delta b_R}_{\rho}.
    \end{aligned} \label{eq:fluctuation_energy_flux}
\end{eqnarray}
In Eq.~(\ref{eq:MHD.dilute_energy_balance}) we have ignored the flux of gas kinetic energy since it appears as a cubic correlation in fluctuating quantities, and in Eqs.~(\ref{eq:averaged_angular_momentum_flux}) and (\ref{eq:MHD.dilute_energy_balance}) we have ignored the Poynting flux ${\bf B}\times\paren{{\bf v}\times{\bf B}}/\paren{4\pi}$ since it is subdominant to other terms in the angular momentum and energy flux. We have ignored terms of the form $\nabla\cdot{\bf v}$ in the viscous stress, which appear in Eqs.~(\ref{eq:averaged_angular_momentum_flux}) and (\ref{eq:MHD.dilute_energy_balance}), since we are in the Boussinesq limit. In the absence of a radiative channel by which the power generated through accretion, $-\paren{\partial\Omega/\partial\ln R}\braket{T_{R\phi}}$, is dissipated or radiated, the energy carried away by the turbulence $\braket{F_{ER}}$ must be positive.

\subsection{Dispersion Relation} \label{sec:dispersion_relation}
Assume small perturbations $\delta a$ about equilibrium fluid quantities $a_0$; thus ${\bf v} = \delta{\bf v}$, ${\bf B} = {\bf B}_0 + \delta{\bf B}$, $\theta = \theta_0 + \delta\theta$, and $\rho = \rho_0 + \delta\rho$. We assume nonaxisymmetric perturbations of the form:
\begin{eqnarray}
  &&\delta a \propto \exp\paren{ik_R R + ik_z z + \gamma t}. \label{eq:perturbed_form}
\end{eqnarray}
In equilibrium the viscous stress and heat flux are zero. One can demonstrate that the perturbed viscous stress tensor and heat flux are given by the following,  where $\delta\bar{\bf B} = \delta{\bf B}/B_0$:
\begin{eqnarray}
  &&\delta\bs{\sigma} = -3\rho_0\nu\paren{{\bf b}{\bf b} - \frac{1}{3}\mathbb{I}}\paren{i\paren{{\bf k}\cdot{\bf b}}\paren{\delta{\bf v}\cdot{\bf b}} + \Omega'R \delta\bar{B}_R \cos\chi}, \label{eq:sigma_perturbed} \\
  &&\delta q = -\kappa p_0\delta\bar{B}_R\parti{\ln T_0}{R} - \kappa p_0\paren{i{\bf k}\cdot{\bf b}}\frac{\delta \theta}{\theta_0}. \label{eq:qv_perturbed}
\end{eqnarray}
The perturbed induction equation, from Eq.~(\ref{eq:induction}) reduces to the following:
\begin{eqnarray}
  &&\begin{aligned}{}
      &\gamma\delta\bar{B}_R = i k_z \sin\chi\delta v_R \\
      &\gamma\delta\bar{B}_{\phi} = i k_z\sin\chi\delta v_{\phi} + \Omega'R\delta\bar{B}_R \\
      &\gamma\delta\bar{B}_z = i k_z B_0\sin\chi\delta v_z,
    \end{aligned} \label{eq:induct}
\end{eqnarray}
From Eq.~(\ref{eq:induct}) the perturbed viscous stress tensor has the following nonzero components:
\begin{eqnarray}
  &&\delta\sigma_{\phi\phi} = \delta\sigma_{zz} = \paren{\cos^2\chi - \frac{1}{3}}\delta\sigma_{{\bf b} {\bf b}}, \label{eq:sigma_phiphi_zz} \\
  &&\delta\sigma_{RR} = -\frac{1}{3}\delta\sigma_{{\bf b} {\bf b}}, \label{eq:sigma_RR} \\
  &&\delta\sigma_{z\phi} = \delta\sigma_{\phi z} = \sin\chi\cos\chi \delta\sigma_{{\bf b} {\bf b}}, \label{eq:sigma_phiz}
\end{eqnarray}
where $\delta\sigma_{{\bf b} {\bf b}}$ is the anisotropic pressure, or equivalently $\delta\sigma_{{\bf b} {\bf b}}/2$ is the component of viscous stress along the magnetic field:
\begin{eqnarray}
  &&\delta\sigma_{{\bf b} {\bf b}} = -3\rho_0\nu \gamma\paren{\delta\bar{B}_{\phi}\cos\chi - \frac{k_R}{k_z}\delta\bar{B}_R\sin\chi}. \label{eq:sigma_Zb_Zb}
\end{eqnarray}
The perturbed energy balance equation, from Eq.~(\ref{eq:energybalance}) is given by the following:
\begin{eqnarray}
  &&\begin{aligned}{}
      &\gamma\paren{\frac{\delta p}{p_0} - \frac{5}{3}\frac{\delta\rho}{\rho_0}} + \delta v_R\parti{\ln p_0\rho_0^{-5/3}}{R} = \\
      &\frac{2}{3}\kappa\paren{ik_z \sin\chi \paren{\delta\bar{B}_R\parti{\ln T_0}{R}} - k_z^2 \sin^2\chi\frac{\delta\theta}{\theta_0}} - \frac{2}{3}p_0^{-1}\delta\bs{\sigma} : \nabla\paren{R\Omega\phivec}.
    \end{aligned} \label{eq:pertenergybalance}
\end{eqnarray}
First, the form of the perturbed stress tensor as given in Eqs.~(\ref{eq:sigma_phiphi_zz}), (\ref{eq:sigma_RR}), and (\ref{eq:sigma_phiz}) implies that $\delta\bs{\sigma} : \nabla\paren{R\Omega\phivec} = 0$. Second, Eq.~(\ref{eq:pertenergybalance}), with isobaric perturbations, implies the following perturbed density:
\begin{eqnarray}
  &&\frac{\delta\rho}{\rho_0} = \frac{3}{5}\times\frac{\delta v_R \partial\ln p_0\rho_0^{-5/3}/\partial R - \frac{2}{3}i\kappa k_z \sin\chi \delta\bar{B}_R\partial\ln T_0/\partial R}{\gamma + \frac{2}{5}\kappa k_z^2\sin^2\chi}. \label{eq:pertrho_simplified}
\end{eqnarray}
Therefore, the perturbed form of the force balance equation,
Eq.~(\ref{eq:forcebalance})
\begin{eqnarray}
  &&\begin{aligned}{}
      \parti{\delta{\bf v}}{t} + 2\Omega\hat\zvec\times\delta{\bf v} + \Omega'R\phivec =& -\frac{1}{\rho_0}\nabla\paren{\delta p + \frac{{\bf b}\cdot\delta{\bf B}}{4\pi}} + \frac{{\bf b}\cdot\nabla\delta{\bf B}}{4\pi\rho_0} - \rho_0^{-1}\nabla\cdot\delta\bs{\sigma} + \\
      &\frac{\delta\rho}{\rho_0}\theta_0\parti{\ln p_0}{R}\Rvec.
    \end{aligned} \label{eq:pertforcebalance}
\end{eqnarray}
In component form this reduces to,
\begin{eqnarray}
  &&\begin{aligned}{}
      \gamma\delta v_R - 2\Omega\delta v_{\phi} =& -ik_R\paren{\frac{\delta p}{\rho_0} + \frac{B_0\cos\chi \delta B_{\phi} + B_0\sin\chi\delta B_z}{4\pi\rho_0}} + \frac{ik_z B_0\sin\chi}{4\pi\rho_0}\delta B_R - \\
      &ik_R\rho_0^{-1}\delta\sigma_{RR} - ik_z\rho_0^{-1}\delta\sigma_{zR} + \frac{\delta\rho}{\rho_0}\theta_0\parti{\ln p_0}{R},
    \end{aligned} \label{eq:pertRforcebalance} \\
  &&\begin{aligned}{}
      \gamma\delta v_{\phi} + \paren{2\Omega + \Omega'R} \delta v_R = \frac{ik_z B_0\sin\chi}{4\pi\rho_0}\delta B_{\phi} - ik_R\rho_0^{-1} \delta\sigma_{\phi R} - ik_z\rho_0^{-1} \delta\sigma_{\phi z},
    \end{aligned} \label{eq:pertphiforcebalance} \\
  &&\begin{aligned}{}
      \gamma\delta v_z =& -ik_z\paren{\frac{\delta p}{\rho_0} + \frac{B_0\cos\chi\delta B_{\phi} + B_0\sin\chi\delta B_z}{4\pi\rho_0}} - \frac{ik_z B_0\sin\chi}{4\pi\rho_0}\delta B_z - \\
      &ik_R \rho_0^{-1}\delta\sigma_{zR} -  ik_z\rho_0^{-1}\delta\sigma_{zz}.
    \end{aligned} \label{eq:pertzforcebalance}
\end{eqnarray}
Eq.~(\ref{eq:pertzforcebalance}) results in the following expression for the total pressure, where $v_A^2 = B_0^2/\paren{4\pi\rho}$:
\begin{eqnarray}
  &&\begin{aligned}{}
      \frac{\delta p}{\rho_0} + v_A^2\cos\chi \delta\bar{B}_{\phi} =& \nu\gamma\paren{3\sin^2\chi - 1}\paren{\delta\bar{B}_{\phi}\cos\chi - \frac{k_R}{k_z} \delta\bar{B}_R\sin\chi} - \frac{ik_R}{k_z^2}\gamma\delta v_R,
    \end{aligned} \label{eq:perturbpressure}
\end{eqnarray}
From the induction equation, Eq.~[\ref{eq:induct}], the force balance equations, Eqs.~(\ref{eq:pertRforcebalance}, \ref{eq:pertphiforcebalance}, \ref{eq:perturbpressure}) can be rewritten in  terms of $\delta\bar{B}_R$ and $\delta\bar{B}_{\phi}$:
\begin{eqnarray}
  &&\begin{aligned}{}
      &\paren{1 + \frac{k_R^2}{k_z^2}}\gamma^2\delta\bar{B}_R - 2\Omega\paren{\gamma\delta\bar{B}_{\phi} - \Omega'R\delta\bar{B}_R} = 3\nu \gamma k_R k_z\sin^3\chi\paren{\delta\bar{B}_{\phi}\cos\chi - \frac{k_R}{k_z}\delta\bar{B}_R\sin\chi} - \\
      &\paren{k_R^2 + k_z^2}v_A^2\sin^2\chi\delta\bar{B}_R + \frac{3}{5}\theta_0\paren{\parti{\ln p_0}{R}}\frac{\gamma\partial\ln p_0\rho_0^{-5/3}/\partial R + \frac{2}{3}\kappa k_z^2 \sin^2\chi \partial\ln T_0/\partial R}{\gamma + \frac{2}{5} \kappa k_z^2 \sin^2\chi}\delta\bar{B}_R,
    \end{aligned} \label{eq:simplifed_radial_force} \\
  &&\begin{aligned}{}
      \gamma^2\delta\bar{B}_{\phi} + 2\Omega\gamma\delta\bar{B}_R = -3\nu \gamma k_z^2 \sin^2\chi\cos\chi\paren{\delta\bar{B}_{\phi}\cos\chi - \frac{k_R}{k_z}\delta\bar{B}_R\sin\chi} - k_z^2 v_A^2\sin^2\chi\delta\bar{B}_{\phi}.
    \end{aligned} \label{eq:equation_two}
\end{eqnarray}
This results in the following dispersion relation:
\begin{eqnarray}
  &&\begin{aligned}{}
      &\lparen{\frac{k^2}{k_z^2}\gamma^2 + \di{\Omega^2}{\ln R} + 3\nu k_R^2\gamma\sin^4\chi - 3\nu k_R k_z\Omega'R \sin^3\chi\cos\chi+ k^2 v_A^2\sin^2\chi -} \\ 
      &\rparen{\frac{3}{5}\theta_0\paren{\parti{\ln p_0}{R}}\frac{\gamma\partial\ln p_0\rho_0^{-5/3}/\partial R + \frac{2}{3}\kappa k_z^2 \sin^2\chi\partial\ln T_0/\partial R}{\gamma + \frac{2}{5}\kappa k_z^2 \sin^2\chi}}\times\\
      &\paren{\gamma^2 + k_z^2 v_A^2\sin^2\chi + 3\nu k_z^2\gamma\sin^2\chi\cos^2\chi} + \gamma^2\paren{4\Omega^2 - 9\nu^2k_R^2k_z^2 \sin^6\chi\cos^2\chi} = 0.
    \end{aligned} \label{eq:totaldisp}
\end{eqnarray}
The Prandtl number for our regime of interest, where electrons dominate the thermal conductivity, is given in \citet{Braginskii65}: 
\begin{eqnarray}
  &&\text{Pr} \equiv \nu/\kappa \approx \frac{0.96}{3.2}\paren{\frac{2m_e}{m_i}}^{1/2} \approx 1/101. \label{eq:Prandtl} 
\end{eqnarray}
After making the following normalizations, where $\alpha_P$, $\alpha_S$, and $\alpha_T$ are the normalized inverse scale heights of pressure, entropy, and temperature, respectively:
\begin{eqnarray}
  &&\begin{aligned}
      &\hat{\bf k} = {\bf k}v_A/\Omega \\
      &\hat{\gamma} = \gamma/\Omega \\
      &\hat{\nu} = \nu\Omega/v_A^2 \\
      &\hat{\kappa} = \kappa\Omega/v_A^2 \\
      &\alpha_P = -H\parti{\ln p_0}{R} \\
      &\alpha_T = -H\parti{\ln T_0}{R} \\
      &\alpha_S = -H\parti{\ln p_0\rho_0^{-5/3}}{R} = \frac{5}{3}\alpha_T - \frac{2}{3}\alpha_P,
    \end{aligned} \label{eq:normalizations}
\end{eqnarray}
where $H = \theta_0^{1/2}/\Omega$ is the disk scale height. One can show that Eq.~(\ref{eq:totaldisp}) reduces to the following:
\begin{eqnarray}
  &&\begin{aligned}{}
      &\lparen{\frac{\hat{k}^2}{\hat{k}_z^2}\hat{\gamma}^2 + 2\di{\ln\Omega}{\ln R} + 3\hat{\nu}\hat{k}_R^2\hat{\gamma}\sin^4\chi - 3\hat{\nu}\hat{k}_R \hat{k}_z \di{\ln\Omega}{\ln R} + \hat{k}^2\sin^2\chi - } \\
      &\rparen{\frac{3}{5}\alpha_P\frac{\alpha_S\hat{\gamma} + \frac{2}{3}\alpha_T\text{Pr}^{-1}\hat{\nu} \hat{k}_z^2\sin^2\chi}{\hat{\gamma} + \frac{2}{5}\text{Pr}^{-1}\hat{\nu}\hat{k}_z^2\sin^2\chi}}\times\paren{\hat{\gamma}^2 + \hat{k}_z^2\sin^2\chi + 3\hat{\nu}\hat{k}_z^2\hat{\gamma}\sin^2\chi\cos^2\chi} + \\
      &\hat{\gamma}^2\paren{4 - 9\hat{\nu}^2\hat{k}_R^2\hat{k}_z^2\sin^6\chi\cos^2\chi} = 0.
    \end{aligned} \label{eq:totdisp_normalized}
\end{eqnarray}
Physically speaking, in this fluid analysis we are always in the regime in which $\kappa > \nu$, hence where $\text{Pr} > 1$. The result is that for sufficiently large transport coefficient $\nu\Omega/v_A^2 > 1$ we then have a density perturbation given by:
\begin{eqnarray*}
  &&\frac{\delta\rho}{\rho_0} \approx \frac{\delta v_R}{\gamma}\paren{\parti{\ln T}{R}},
\end{eqnarray*}
and the following substitution occurs in Eq.~(\ref{eq:totdisp_normalized}):
\begin{eqnarray*}
  &&\frac{3}{5}\alpha_P\frac{\alpha_S\hat{\gamma} + \frac{2}{3}\alpha_T\text{Pr}^{-1}\hat{\nu} \hat{k}_z^2\sin^2\chi}{\hat{\gamma} + \frac{2}{5}\text{Pr}^{-1}\hat{\nu}\hat{k}_z^2\sin^2\chi} \to \alpha_P \alpha_T.
\end{eqnarray*}
Note the limits of Eq.~(\ref{eq:totdisp_normalized}): 1) in the limit that the transport coefficients go to zero and there are no equilibrium gradients, we reproduce the MRI; 2) in the limit of no equilibrium gradients and finite viscosity, we reproduce the MVI; 3) in the limit of no viscosity but finite thermal conductivity, we reproduce the MTI dispersion relation \citep{Balbus01a}; and 4) in the limit of no thermal conductivity or viscosity, but finite equilibrium gradients, we reproduce the dispersion relation for convectively unstable modes in a rotating magnetized plasma  \citep{Balbus95}. To put these parameters in perspective, in physical terms the condition under which the magnetoviscous instability operates is one in which $\nu\Omega/v_A^2 > 1$. The viscosity and Alfv\'{e}n velocity in dimensional units is given by the following:
\begin{eqnarray}
  &&\nu = 1.4\times 10^{19} \frac{T_4^{5/2}}{n_1\ln\Lambda}\text{ cm$^2$ s$^{-1}$}, \label{eq:nu_dimensional} \\
  &&v_A = 2.2\times 10^5 B_{\mu G} n_1^{-1/2}\text{ cm s$^{-1}$}. \label{eq:vA_dimensional}
\end{eqnarray}
One must thus have an angular velocity of the flow such that:
\begin{eqnarray}
  &&\Omega \gtsim 3.5\times 10^{-9} B_{\mu G}^2\paren{\ln\Lambda} T_4^{-5/2}\text{ s$^{-1}$},
\end{eqnarray}
for the viscosity to be dynamically important. Furthermore, the thermal diffusivity $\kappa \gg \nu$, so that the range of applicability of thermal effects is significantly larger than that of viscous transport.

\section{Results} \label{sec:results}
In this section we calculate the quadratic azimuthal stress $\braket{T_{R\phi}}$ and radial heat flux $q_R$ associated with these instabilities. Since the MVTI operates in a regime of high thermal diffusivity and possibly high viscous diffusivity, we need to consider the local energetics and angular momentum transport modified by large viscous and thermal fluxes. From Eqs.~(\ref{eq:averaged_angular_momentum_flux}) and (\ref{eq:fluctuation_energy_flux}), we estimate the quadratic angular momentum flux and heat flux carried away by a mode of wavenumber ${\bf k}$ as the following,
\begin{eqnarray}
  &&\begin{aligned}
      &\braket{T_{R\phi}} = \text{Re}\lparen{\rho_0 \delta u_{R} \delta u_{\phi}^* - \rho_0 v_A^2 \delta\bar{B}_{R} \delta\bar{B}_{\phi}^* -} \\
      &\rparen{3i\rho_0 \nu\brak{\paren{{\bf k}\cdot{\bf b}_0}\paren{\delta{\bf v}_{\bf k}\cdot{\bf b}_0} + \Omega'R \delta\bar{B}_{R}\cos\chi} \delta\bar{B}_{R}^*\cos\chi},
    \end{aligned} \label{eq:angular_momentum_flux_modal} \\
  &&\begin{aligned}{}
      &\braket{F_{ER}} = \text{Re}\lparen{\frac{5}{2}\rho_0 \delta u_{R} \delta\theta_{\bf k}^* - \kappa \rho_0 \parti{\theta_0}{R} \abs{\delta\bar{B}_{R}}^2 + }\\
      &\rparen{i\rho_0 \nu\brak{\paren{{\bf b}_0\cdot\delta{\bf v}_{\bf k}}\paren{{\bf b}_0\cdot{\bf k}} + \Omega'R \delta\bar{B}_{R}\cos\chi}\delta\bar{B}_{R}^*\rule{0em}{1.5em}}.
    \end{aligned} \label{eq:heat_flux_modal}
\end{eqnarray}
$\delta a$ refers to fluctuating quantity $a$ for an MVTI eigenmode of wavenumber ${\bf k}$.

\subsection{Growth Rate and Stability Characteristics}
\begin{figure}[!ht]
  \includegraphics[width=\linewidth]{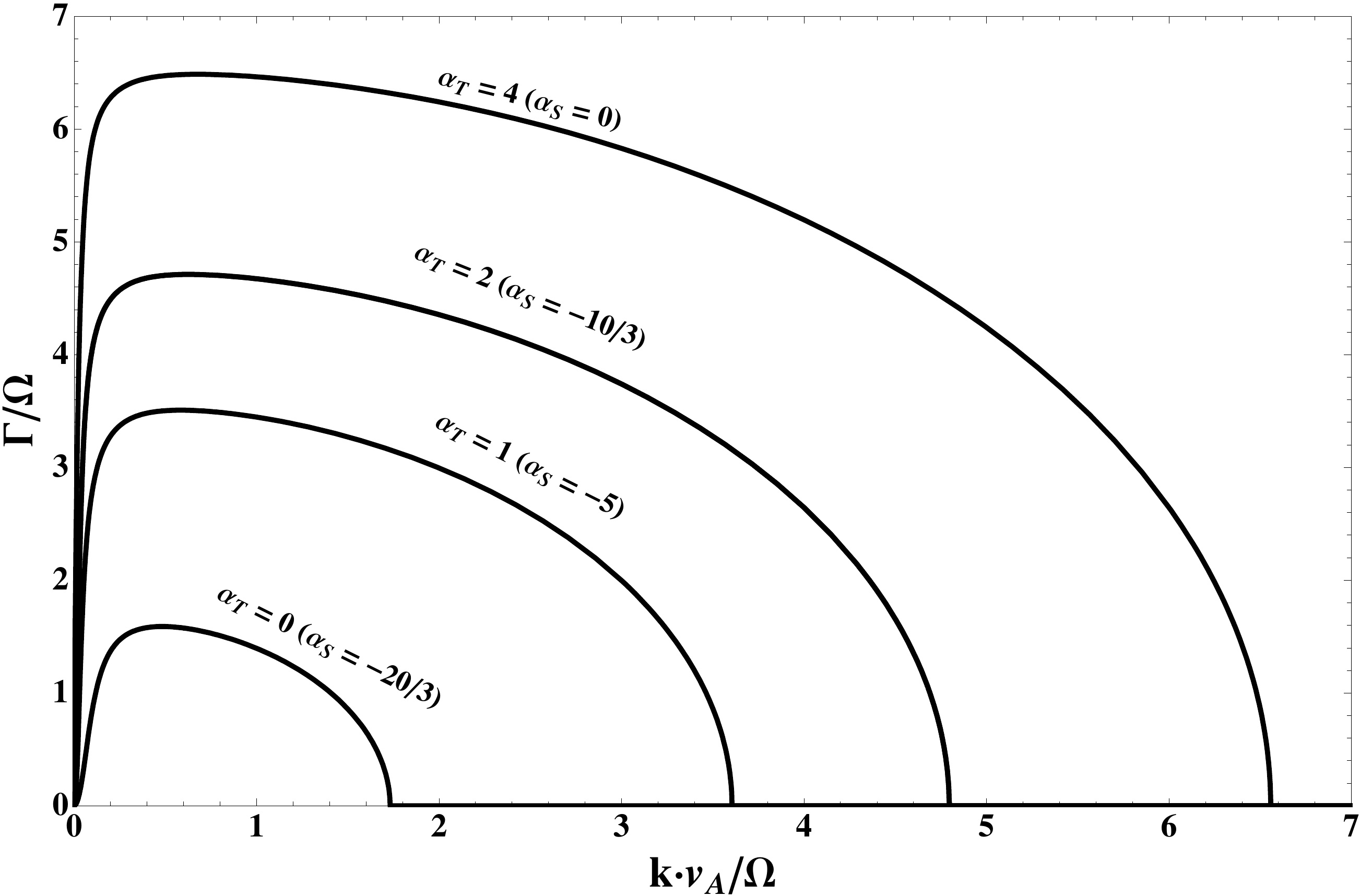}
  \caption{Plot of the real portion of the growth rate for various $\alpha_T$ for a Keplerian-like rotation profile, $\Omega\propto R^{-3/2}$, significant viscous diffusion coefficient $\nu\Omega/v_A^2 = 10^2$ and Prandtl number $\text{Pr} = 1/101$ (see Eq.~[\ref{eq:Prandtl}]) -- this corresponds to the case where both anisotropic viscous forces dominate over magnetic tension.} \label{fig:mvi}
\end{figure}

\begin{figure}[!ht]
  \includegraphics[width=\linewidth]{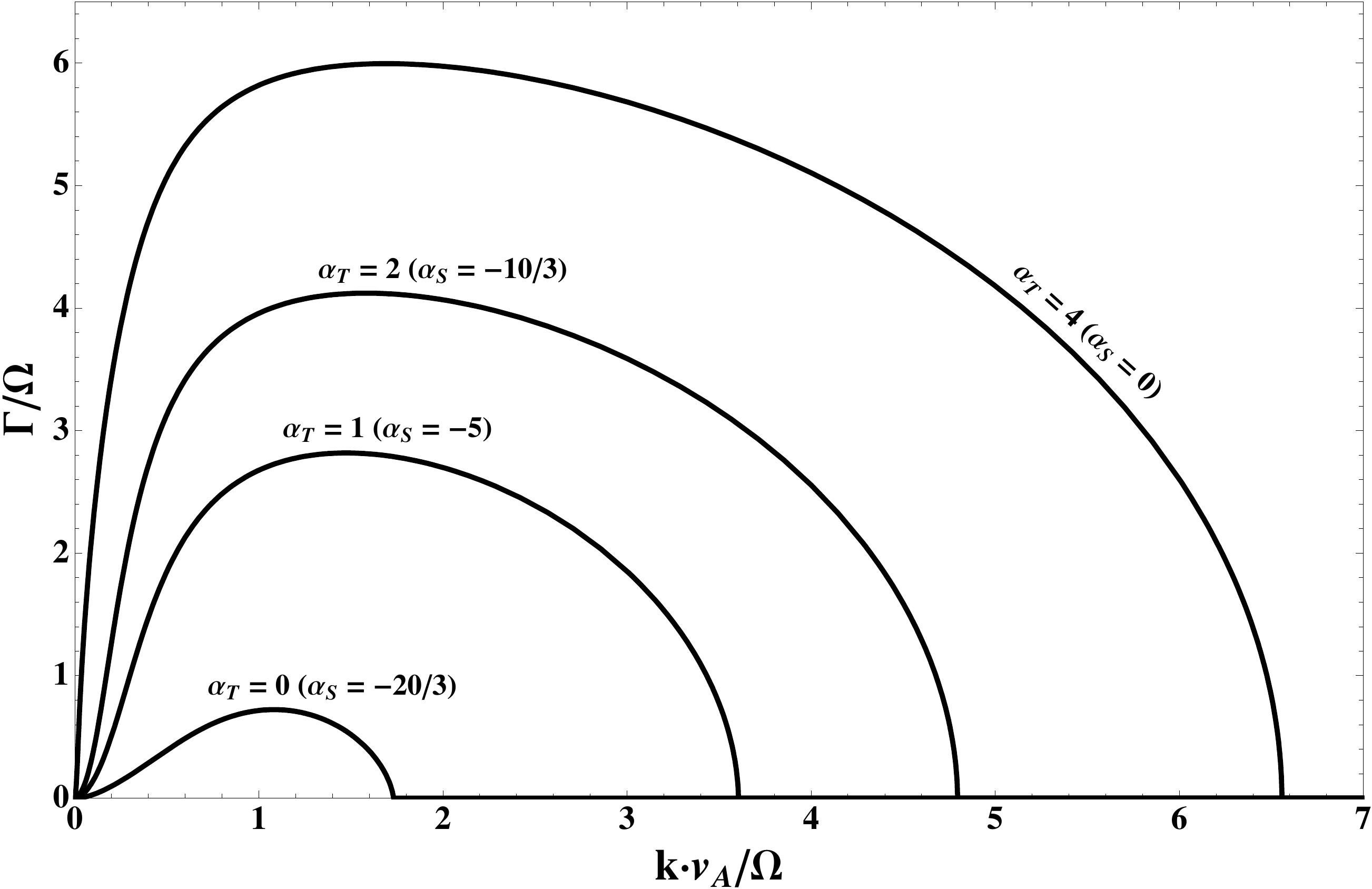}
  \caption{Plot of the dispersion relation for various $\alpha_T$ for a Keplerian-like rotation profile, $\Omega\propto R^{-3/2}$, a small viscous diffusion coefficient $\nu\Omega/v_A^2 = 1$ and Prandtl number $\text{Pr} = 1/101$ (see Eq.~[\ref{eq:Prandtl}]).} \label{fig:mri}
\end{figure}

Here we consider two equilibrium configurations that illustrate the magnetoviscous-thermal dispersion relation. We only consider a physical pressure profile, hence $\alpha_P = 10$ (i.e., outwardly decreasing pressure). We also consider a flow that is convectively stable, hence with $H\partial\ln P\rho^{-5/3}/\partial R > 0$ or $\alpha_S < 0$. This implies the following:
\begin{eqnarray}
  &&0 < \alpha_T \le \frac{2}{5}\alpha_P, \label{eq:Schwarzchild_condition}
\end{eqnarray}
which implies that $\alpha_T \le 4$. We focus our results on only vertical wavenumbers, $k_R = 0$. Then Eq.~(\ref{eq:totdisp_normalized}) reduces to the following quintic polynomial, where we denote $x = \paren{{\bf k}\cdot{\bf v}_A}/\Omega = \hat{k}_z \sin\chi$:
\begin{eqnarray}
  &&\begin{aligned}{}
      &\paren{\brak{\hat{\gamma}^2 + 2\di{\ln\Omega}{\ln R} + x^2}\brak{\hat{\gamma} + \frac{2}{5}\text{Pr}^{-1}\hat{\nu} x^2} - \frac{3}{5}\alpha_P\brak{\alpha_S\hat{\gamma} + \frac{2}{3}\alpha_T\text{Pr}^{-1}\hat{\nu} x^2}}\times \\
      &\paren{\hat{\gamma}^2 + x^2 + 3\hat{\nu}\hat{\gamma} x^2\cos^2\chi} + 4\hat{\gamma}^2\paren{\hat{\gamma} + \frac{2}{5}\text{Pr}^{-1}\hat{\nu} x^2} = 0,
    \end{aligned} \label{eq:totdisp_normalized_simplified}
\end{eqnarray}
In Fig.~(\ref{fig:mvi}), we examine the instance where both magnetic tension and thermal conductivity are dynamically significant; dispersion relations match the salient characteristic of the magnetoviscous instability \citep{Balbus04b, Islam2005} -- saturation of the growth rate at small wavenumbers, described by the condition $\nu k^2 \sim \Omega$. For the case $\alpha_T = 0$ (no equilibrium temperature gradient) we approach the magnetoviscous instability -- the mode is reproduced only for $\alpha_P = 0$ and $\alpha_T = 0$. In Fig.~(\ref{fig:mri}), we examine the instance where only the thermal conductivity is dynamically important; the dispersion relations are similar to the magnetorotational instability. From an examination of the dispersion relation Eq.~(\ref{eq:totaldisp}) or the 	normalized dispersion relation  Eq.~(\ref{eq:totdisp_normalized_simplified}) that for physical $\alpha_T > 0$ the growth rate and wavenumber extent of the instability are increased -- this is shown also in Fig.~(\ref{fig:mvi}) and Fig.~(\ref{fig:mri}).
One expects that equilibrium scale heights of temperature, pressure, and entropy are of order the radius $R$. Thus, in order to have significant magnetoviscous and magnetothermal effects one requires that $\alpha_T$, $\alpha_P$, and $\alpha_S$ be at least of order unity. Hence, only thick disks $H \sim R$ are expected to be susceptible to these classes of instability. However, note that we have calculated quadratic modal fluxes of ostensibly thick disks using thin-disk expressions for angular momentum (Eq.~[\ref{eq:angular_momentum_flux_modal}]) and heat (Eq.~[\ref{eq:heat_flux_modal}]). This may not pose an intractable problem: first, even for relatively slender disks, ones in which $H \ltsim R$, we can expect that the angular momentum and heat fluxes due to the nonlinear MVTI will still be of the same sign (outwards) and magnitude as their respective thin-disk expressions; and second, even for thin RIAFs, the nonlinear MVTI or its collisionless analogue appear to provide the only plausible MHD mechanism for the transport of accretion energy outwards.

\subsection{Heat Flux and Azimuthal Stresses} \label{sec:quadfluxes}
\begin{figure}[!ht]
  \includegraphics[width=\linewidth]{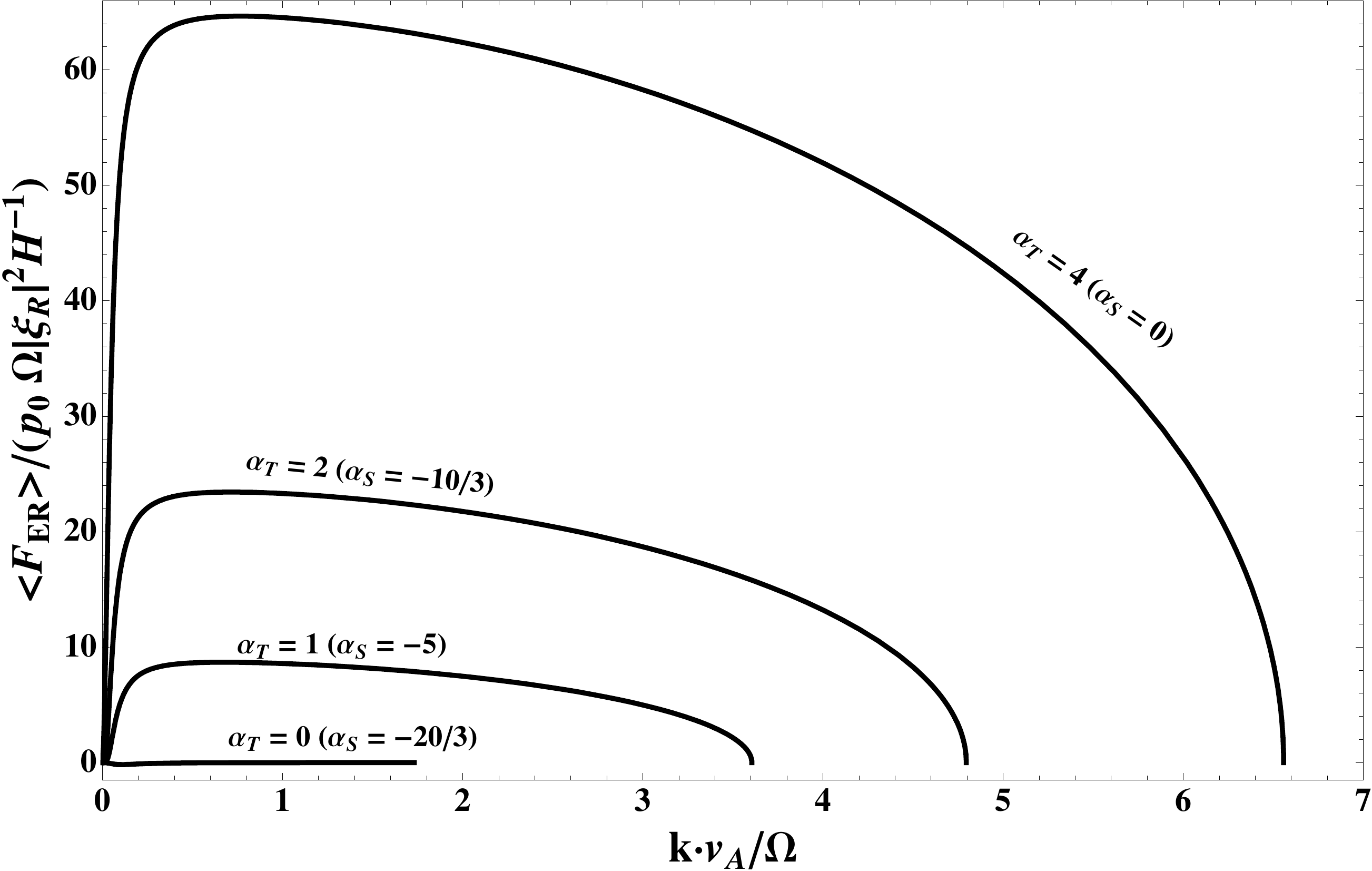}
  \caption{The normalized radial flux of thermal energy with parameters as described in Fig.~(\ref{fig:mvi})), where $\nu\Omega/v_A^2 = 10^2$.} \label{fig:mvi_HeatFlux}
\end{figure}

\begin{figure}[!ht]
  \includegraphics[width=\linewidth]{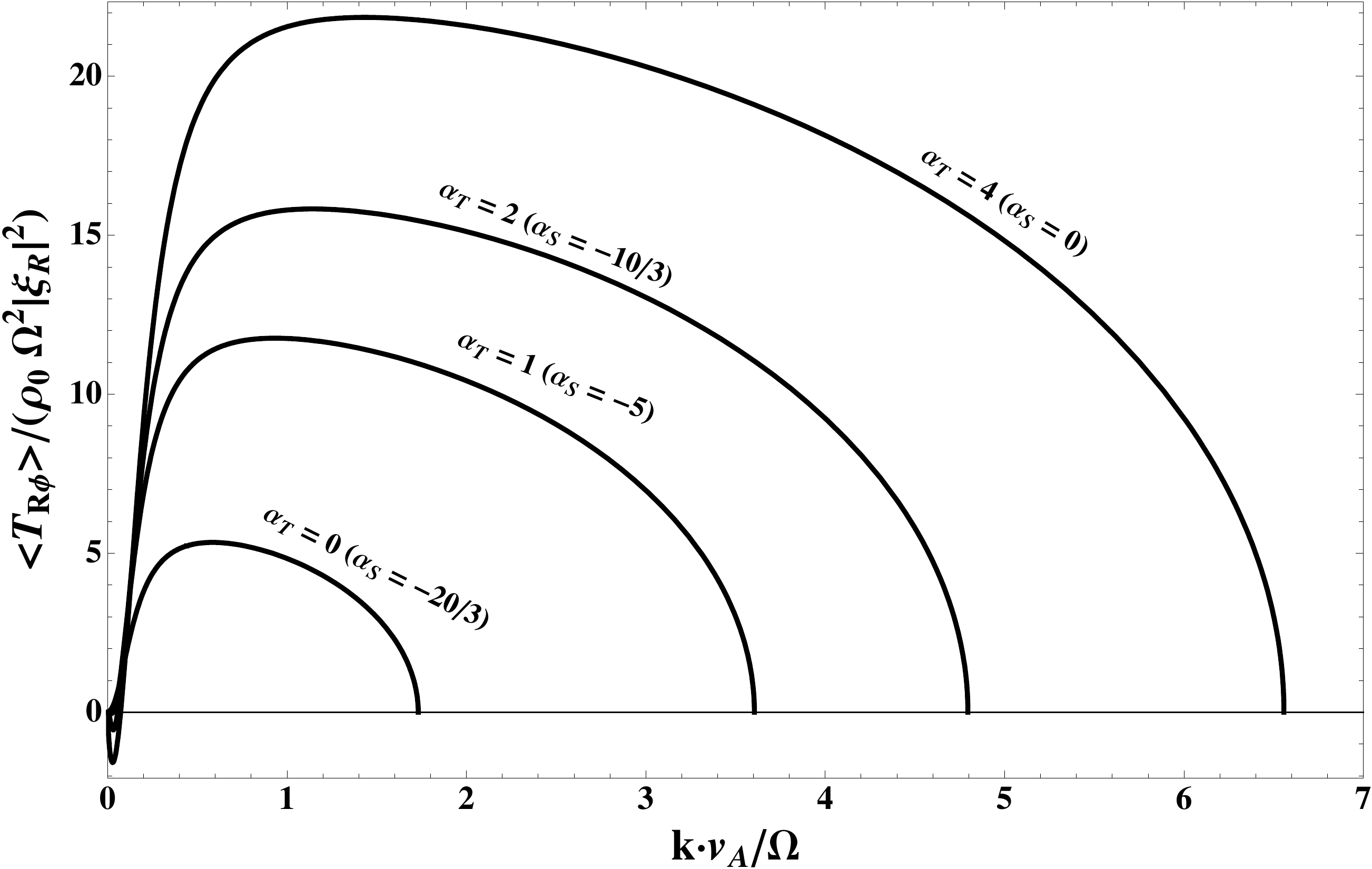}
  \caption{Normalized azimuthal momentum radial flux with parameters as described in Fig.~(\ref{fig:mvi}), where $\nu\Omega/v_A^2 = 10^2$.} \label{fig:mvi_AngmomFlux}
\end{figure}

\begin{figure}
  \includegraphics[width=\linewidth]{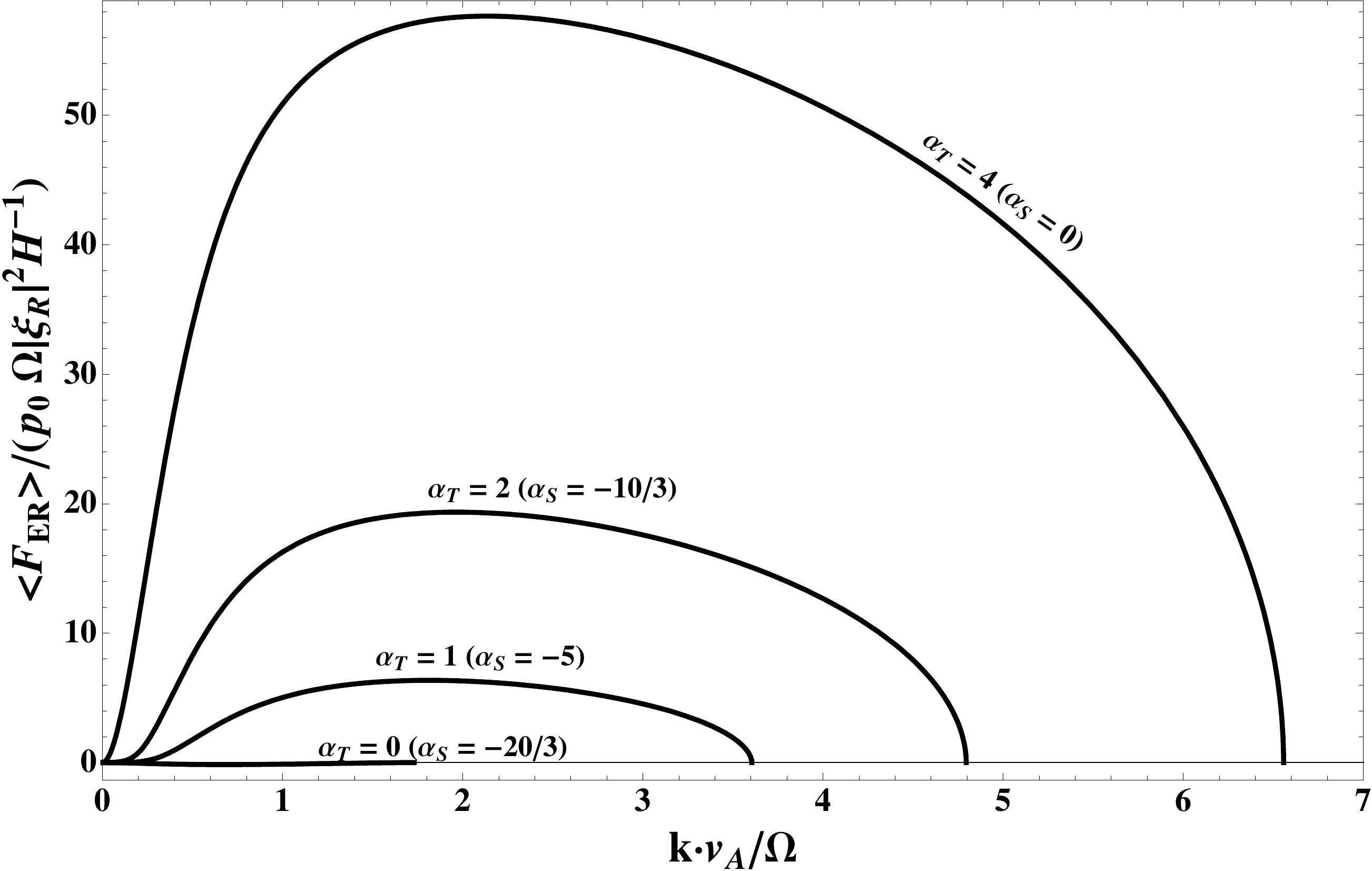}
  \caption{The normalized radial flux of thermal energy for a magnetorotational-like mode, with parameters as described in Fig.~(\ref{fig:mri}).} \label{fig:mri_HeatFlux}
\end{figure}

\begin{figure}
  \includegraphics[width=\linewidth]{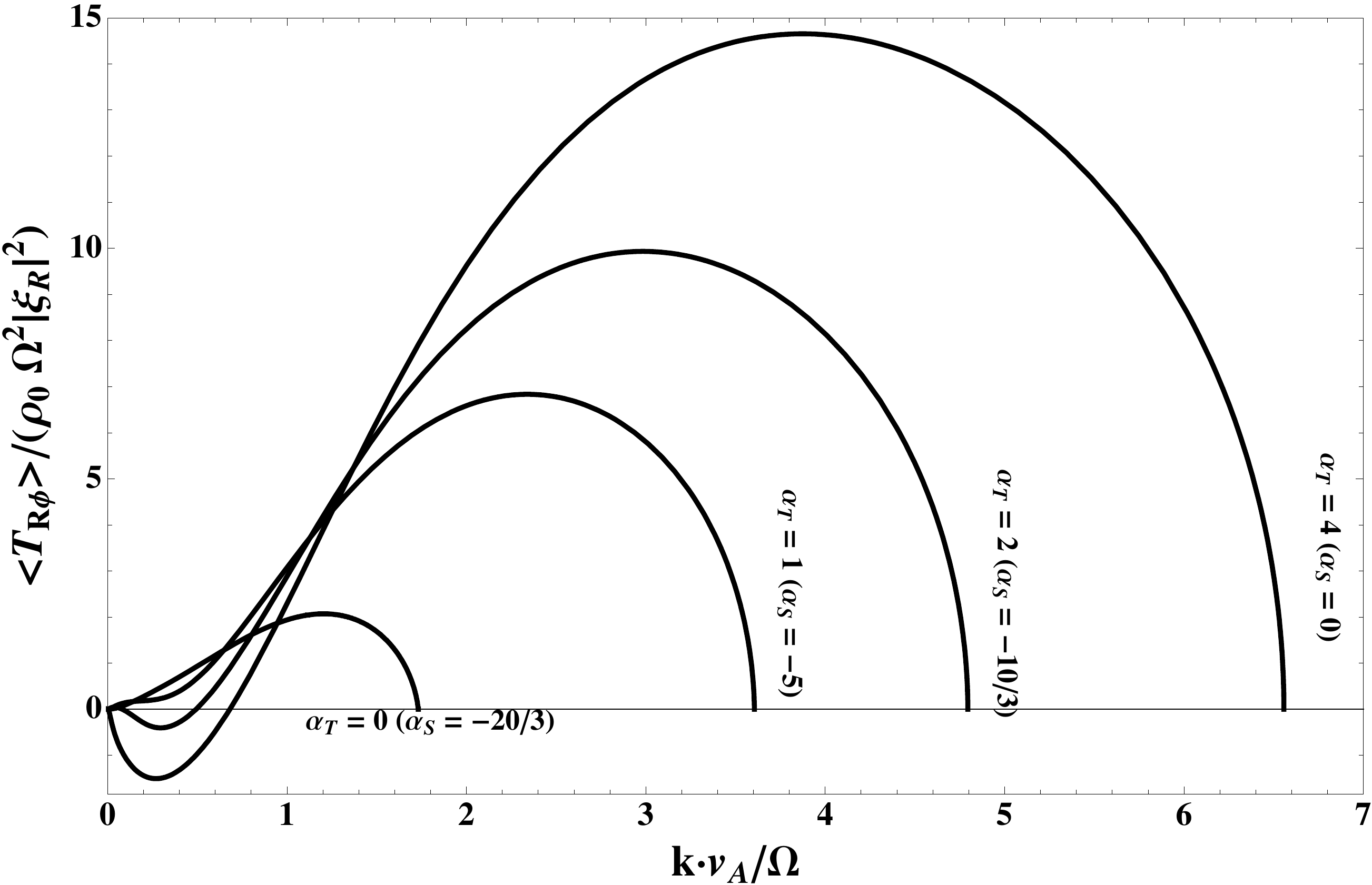}
  \caption{Normalized $\braket{T_{R\phi}}$ for a magnetorotational-like mode, with parameters as described in Figs.~(\ref{fig:mri}). For magnetoviscous-like modes, it is apparent that there is a range of low wavenumbers for which $\braket{T_{R\phi}}$ is inwards.} \label{fig:mri_AngmomFlux}
\end{figure}

The following are the calculated perturbed velocities, magnetic
fields, density, and temperature in terms of the radial displacement
$\xi_R = \gamma^{-1} \delta v_R$, that employing the induction
equations (Eq.~[\ref{eq:induct}]) and
Eqs.~(\ref{eq:pertrho_simplified}) and
(\ref{eq:equation_two}). We examine nonradial modes, $k_R/k_z = 0$:
\begin{eqnarray}
  &&\begin{aligned}{}
      &\delta v_R = \hat{\gamma} \Omega\xi_R \\
      &\delta v_{\phi} = -\paren{\frac{2\hat{\gamma}^2}{\hat{\gamma}^2 + x^2 + 3\hat{\nu} x^2 \hat{\gamma} \cos^2\chi} + \di{\ln\Omega}{\ln R}}\Omega\xi_R \\
      &\delta B_R = i\paren{{\bf k}\cdot{\bf B}}\xi_R \\
      &\delta B_{\phi} = -i\paren{{\bf k}\cdot{\bf B}}\frac{2\hat{\gamma}}{\hat{\gamma}^2 + x^2 + 3\hat{\nu} x^2 \hat{\gamma}\cos^2\chi}\xi_R \\ 
      &\delta\rho = -\frac{3}{5}\times\frac{\hat{\gamma}\alpha_S + \frac{2}{3}\text{Pr}^{-1}\hat{\nu} x^2 \alpha_T}{\hat{\gamma} + \frac{2}{5}\text{Pr}^{-1}\hat{\nu} x^2}\paren{\rho_0 H^{-1}\xi_R}, \\
      &\delta\theta = -\theta_0 \delta\rho/\rho_0 = \frac{3}{5}\times\frac{\hat{\gamma}\alpha_S + \frac{2}{3}\text{Pr}^{-1}\hat{\nu} x^2 \alpha_T}{\hat{\gamma} + \frac{2}{5}\text{Pr}^{-1}\hat{\nu} x^2}\paren{\theta_0 H^{-1}\xi_R},
    \end{aligned} \label{eq:perturbed_quantities}
\end{eqnarray}
with the form of the perturbed anisotropic pressure from Eq.~(\ref{eq:sigma_Zb_Zb}), the turbulent heat flux from Eq.~(\ref{eq:heat_flux_modal}) yields the following: 
\begin{eqnarray}
  &&\begin{aligned}{}
      F_{ER} &= \rho_0\text{Re}\paren{\frac{5}{2}\delta\theta\delta v_R^{*} + \rho_0\nu\gamma \cos\chi \delta v_R^* \delta\bar{B}_{\phi}} - \rho_0 \text{Pr}^{-1} \nu k_z \text{Im}\paren{\sin\chi \delta\theta\delta\bar{B}_R^*} \\
      &=\paren{\frac{3}{2}\hat{\gamma}\frac{\hat{\gamma}\alpha_S + \frac{2}{3}\hat{\nu}\text{Pr}^{-1}x^2\alpha_T}{\hat{\gamma} + \frac{2}{5}\hat{\nu}\text{Pr}^{-1}x^2}}p_0\Omega\abs{\xi_R}^2 H^{-1}, \end{aligned}
  \label{eq:radialthermalflux}
\end{eqnarray}
And the azimuthal stress from Eq.~(\ref{eq:angular_momentum_flux_modal}) yields the following:
\begin{eqnarray}
  &&\begin{aligned}{}
      T_{R\phi} &= \text{Re}\paren{\rho_0\delta v_R\delta v_{\phi}^{*} - \frac{\delta B_R\delta B_{\phi}^{*}}{4\pi} - 3\rho_0 \nu \cos^2\chi\gamma\delta\bar{B}_R^* \delta\bar{B}_{\phi}} \\
      &=\hat{\gamma}\paren{2 - \di{\ln\Omega}{\ln R} - \frac{4\hat{\gamma}^2}{\hat{\gamma}^2 + x^2 + 3\hat{\nu}x^2\hat{\gamma}\cos^2\chi}}\rho_0\Omega^2\abs{\xi_R}^2,
    \end{aligned} \label{eq:angmomflux}
\end{eqnarray}
In all the quasilinear plots we keep $\alpha_P = 10$, choose $\alpha_T > 0$ for which the flow is convectively stable, and fix the Prandtl number $\text{Pr} = 1/101$. In Figs.~(\ref{fig:mvi_HeatFlux}) and (\ref{fig:mvi_AngmomFlux}) are plots of the normalized thermal flux $q_R$ and azimuthal stress $T_{R\phi}$ for magnetoviscous-like dispersion modes ($\nu \Omega/v_A^2 = 10^2$, see Fig.~[\ref{fig:mvi}]). In Figs.~(\ref{fig:mri_HeatFlux}) and (\ref{fig:mri_AngmomFlux}) are plots of normalized $q_R$ and $T_{R\phi}$ for magnetorotational-like modes ($\nu\Omega/v_A^2 = 1$,
see Fig.~[\ref{fig:mri}]).
From Figs.~(\ref{fig:mvi_HeatFlux}) and (\ref{fig:mri_HeatFlux}), one observes that the quadratic heat flux in the limit of large ($\nu\Omega/v_A^2 \gg 1$) and moderate ($\nu\Omega/v_A^2 \sim 1$) viscosities results in dynamically significant outwards heat fluxes, with $q_R \gtsim 	T_{R\phi}\paren{k_B\theta_0/m_i}^{1/2}$. This implies that the nonlinear MVTI can play an important role in transporting out the energy, generated via the azimuthal stress, in nonradiative  accreting flows. Second, from Figs.~(\ref{fig:mvi_AngmomFlux}) andm (\ref{fig:mri_AngmomFlux}), one notes that the flux of angular momentum for these modes can be either inwards or outwards. One also observes for a range of small wavenumbers that thermal gradients may cause a negative azimuthal stress, unlike the MRI.

\begin{figure}[!ht]
  \includegraphics[width=\linewidth]{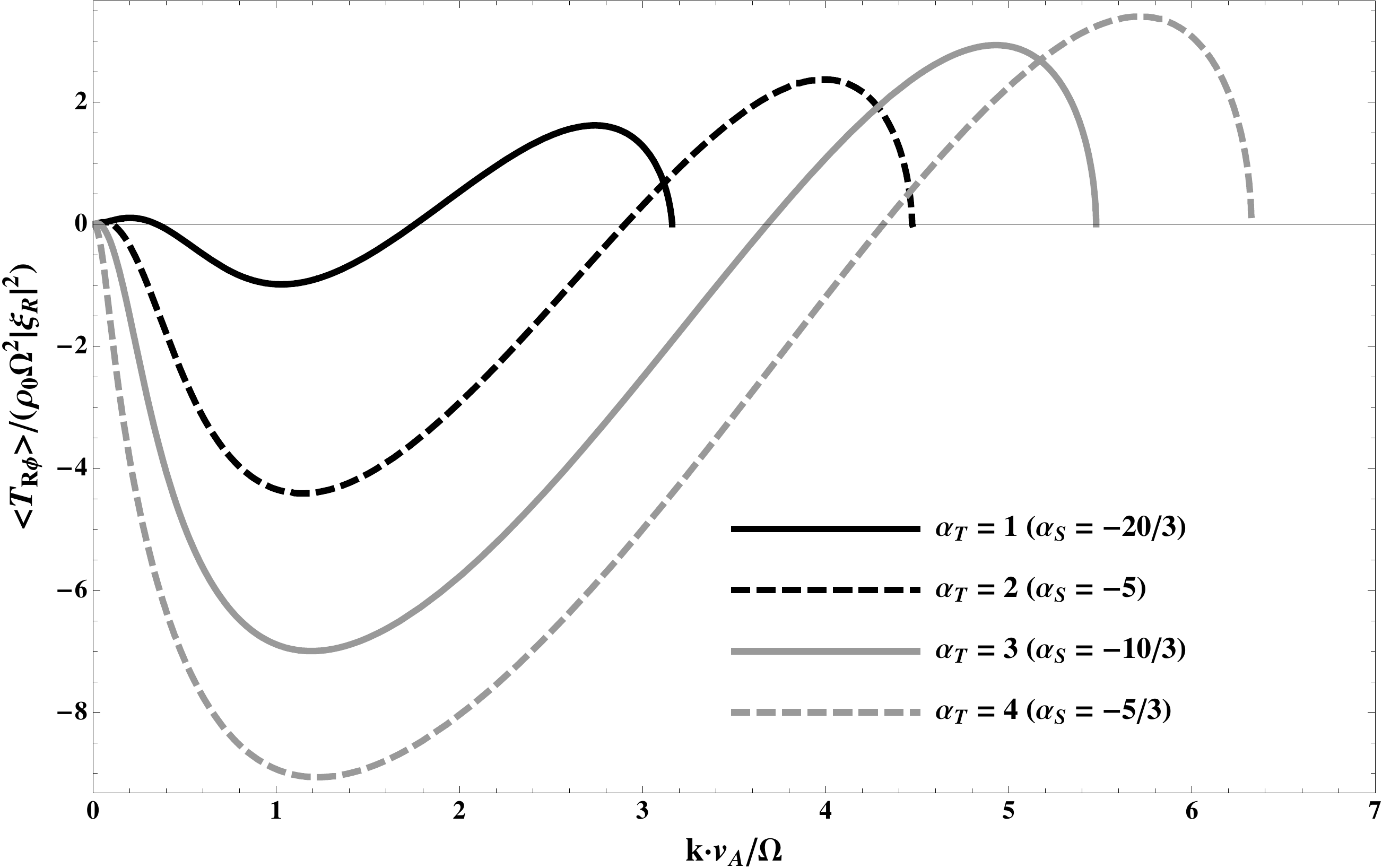}
  \caption{Normalized flux $\braket{T_{R\phi}}/\paren{\rho\Omega^2\abs{\xi_R}^2}$ for a rigid rotation profile ($\Omega'R = 0$) for the magnetothermal instability (see Eq.~[\ref{eq:MTI_dispersion}]). There is a much larger range of negative $\braket{T_{R\phi}}$ (inwards flux of angular momentum) than shown in Fig.~(\ref{fig:mri_AngmomFlux}).} \label{fig:MTI_angmomflux_rigidrotation}
\end{figure}

An important point that must be made is that with a large thermal conductivity, angular momentum momentum may be transported inwards or outwards for large decreasing gradients of temperature. This effect can be seen even for the purely magnetothermal instability (MTI) in a rigidly rotating plasma. We can re-derive the dispersion relation and angular momentum flux for the MTI by setting $\text{Pr}^{-1} = \kappa/\nu$, while letting $\nu \to 0$. Then the normalized dispersion relation, Eq.~(\ref{eq:totdisp_normalized_simplified}), and azimuthal stress, Eq.~(\ref{eq:angmomflux}):
\begin{eqnarray}
  &&\begin{aligned}{}
      &\paren{\brak{\hat{\gamma}^2 + 2\di{\ln\Omega}{\ln R} + x^2}\brak{\hat{\gamma} + \frac{2}{5}\hat{\kappa} x^2} - \frac{3}{5}\alpha_P\brak{\alpha_S\hat{\gamma} + \frac{2}{3}\alpha_T\hat{\kappa} x^2}}\paren{\hat{\gamma}^2 + x^2} + \\
      &4\hat{\gamma}^2\paren{\hat{\gamma} + \frac{2}{5}\hat{\kappa} x^2} = 0,
    \end{aligned} \label{eq:MTI_dispersion} \\
  &&T_{R\phi} = \hat{\gamma}\frac{x^2\paren{2 - d\ln\Omega/d\ln R} - \hat{\gamma}^2\paren{2 + d\ln\Omega/d\ln R}}{\hat{\gamma}^2 + x^2}\rho\Omega^2\abs{\xi_R}^2. \label{eq:MTI_angmomflux}
\end{eqnarray}
Fig.~(\ref{fig:MTI_angmomflux_rigidrotation}) demonstrates that in a rigidly rotating plasma $\Omega'R = 0$, the magnetothermal instability can drive a positive azimuthal stress depending on wavenumber. As the system approaches marginal convective stability $\alpha_S \to 0$ from isothermality ($\alpha_T = 0$), the range of wavenumbers for which the stress is outwards decreases. One must note, however, that in the absence of rotational shear no energy can be extracted from the flow. Second, the ambiguity of angular momentum transport for the MVTI in this case is not necessarily evidence of unusual physics: the MVTI acts as a mechanism to transport thermal energy outwards, which is largely (and, in the case of rigid rotation, completely) independent of the manner in which it transports angular momentum. A first step into a numerical study to characterize MVTI stability could begin with a high spatial grid resolution simulation, at especially large Reynolds numbers, about a near-MVTI stable rotating system, in order to discover unambiguous (not numerical diffusivity limited) levels of turbulence and angular momentum and heat flux. This would be analogous to \citet{Lesur05}, who have demonstrated the unambiguous presence of minuscule turbulence in high Reynolds number (Re $\sim 10^5$) hydrodynamic numerical simulations of differentially rotating flows near Rayleigh stability.
\FloatBarrier

\section{Conclusions and Further Research}
In this paper we attempted to characterize the stability of a mildly dilute rotating accretion flow to axisymmetric magnetoviscous-thermal modes, by incorporating both an anisotropic magnetized viscous stress tensor and anisotropic electron thermal conductivities. We have demonstrated several important properties of the MVTI. First, we find that thick disks with sufficiently large thermal diffusion coefficients are susceptible to these classes of temperature gradient instabilities. Second, physical outwardly decreasing temperature gradients increase the range of wavenumbers for which the magnetorotational-like instability (where the viscous diffusion coefficient is small $\nu\Omega < v_A^2$) and magnetoviscous-like instability (where the viscous diffusion coefficient is large) are unstable, as well as increasing the growth rate of unstable wavenumbers. Third, quadratic fluxes of angular momentum are substantially modified by both a large outward viscous transport in dilute plasmas. Fourth, in rotating systems with equilibrium outwardly decreasing temperature, the linear MVTI can transport angular momentum either inwards or outwards.

Companion papers will consider the following. First, we will analyze a more general case of the collisionless and mildly collisional MTI, demonstrating that there is an effective viscous and thermal diffusivity of order $\theta_0/\Omega$ along the magnetic field lines. Second, we demonstrate the dynamics of the collisionless and mildly collisional MTI with electron pressure dynamics, in order to characterize the level of electron (and ion) heating due to nonlinear collisionless MTI turbulence. And third, we consider more unusual physical phenomena (finite equilibrium acceleration of ions and electrons away from the disk midplane, or collisionless reconnecting modes in a Keplerian profile) in collisionless disks. A necessary next step would be to consider more carefully an appropriate local equilibrium profile (see, e.g., \citet{Bisnovatyi-Kogan72, Bisnovatyi-Kogan85}).

Our analysis of the MVTI may allow for a clearer physical resolution of analytic models, such as proposed advection dominated accretion flows \citep{Narayan94}, convectively dominated accretion flows \citep{Quataert00}, or adiabatic inflow-outflow solutions \citep{Blandford99}, of largely rotationally supported RIAFs. Energy transport for Rayleigh-stable flows is generally outwards where $\paren{\partial\ln T/\partial\ln R} > 0$, and the maximal rate of quadratic energy transport $F_{ER}$ goes as $H/R\paren{\partial\ln T/\partial\ln R}$. This implies that the level of MHD turbulence and the level of energy transport in RIAFs may be some function of this parameter. In analytic models of accretion in dilute rotationally supported flows, a more appropriate estimate of the turbulent energy transport flux may be $H/R\paren{\partial \ln T/\partial\ln R} p \theta^{1/2}$; this is in contrast to an energy flux estimate of $p\theta^{1/2}$, applicable to largely pressure supported or spherically symmetric flows \citep{Tanaka06, Johnson07}. For geometrically thin RIAFs, due to quadratic estimates dependent on the parameter $H \partial\ln T/\partial R$, our results suggest that accretion will be far below the gravitational capture rate of matter from the black hole's ambient medium.

Although much work remains to be done in understanding the dynamics of hot dilute accretion flows, this paper and its companions \citep{Balbus01a, Balbus04b, Islam2005} have demonstrated a class of instabilities that may roughly describe the type of turbulence expected within RIAFs. Unfortunately, MHD simulations of a local radial slice of a Keplerian disk are ill-suited to studies of phenomena in which both angular momentum and energy transport must be characterized, as the total energy within the radial slice increases at a rate $-\paren{\partial\Omega/\partial\ln R}\braket{T_{R\phi}}$. Global collisionless MHD simulations, employing fluid codes that naturally conserve total mechanical energy such as ATHENA \citep{Stone08}, show the most promise in beginning our understanding of the nonlinear stages of these newly explored free-energy channels, and possibly in demonstrating accretion via nonlinear MVTI turbulence.

\section{Acknowledgments}
The author acknowledges the generous support of Steven Balbus, in introducing the author to novel mechanisms for accretion in radiatively inefficient accretion flows, and in his support of this research at the University of Virginia and the \'{E}cole Normale Superieure. The author also wishes to acknowledge Craig Markwardt, for his explanation of recent SWIFT BAT X-ray observational evidence of cosmologically local active galactic nuclei.
\bibliography{apj-jour,fullbib}
\end{document}